\documentclass[final,3p,times,twocolumn]{elsarticle}

\usepackage{flushend}
\usepackage{times}
\usepackage{helvet}
\usepackage{courier}
\usepackage{wrapfig}

\usepackage{graphicx}
\usepackage{amssymb,amsmath,bm}
\usepackage{textcomp}
\usepackage{flushend}
\usepackage{subfigure}
\usepackage{url}
\usepackage{array, multirow}
\usepackage{threeparttable}
\usepackage{threeparttable}
\usepackage{booktabs} %
\usepackage{lettrine}

\PassOptionsToPackage{dvipsnames}{xcolor}  
\usepackage{tikz}
\usetikzlibrary{arrows,positioning,fit,petri,shapes,backgrounds,decorations.pathmorphing,calc,shapes.misc, arrows, decorations.markings}

\newcommand{\eg}{e.\,g., }
\newcommand{\ie}{i.\,e., }

\newcommand{\mb}{\mathbf}

\usepackage{xcolor}
\definecolor{newgreen}{RGB}{34,139,34}
\newcommand{\re}[1] {{\color{newgreen} \bf #1}} 

\usepackage{soul}

\journal{*}

\begin{document}

\begin{frontmatter}


\title{Learning Audio Sequence Representations for Acoustic Event Classification}



\author{Zixing Zhang$^{1}$, Ding Liu$^{2}$, Jing Han$^{3}$, Kun Qian$^{4*}$, and Bj\"orn W.\ Schuller$^{1,3}$}

\address{1.\,GLAM -- Group on Language, Audio \& Music, Imperial College London, London SW7 2AZ, UK,\\E-mail:~connectzzx@gmail.com, bjoern.schuller@imperial.ac.uk \\
         2.\,Department of Informatics, Technische Universit\"at M\"unchen, 80333 Munich, Germany, E-mail:~ding.liu@tum.de \\
         3.\,ZD.B Chair of Embedded Intelligence for Health Care and Wellbeing, University of Augsburg, Augsburg 86159, Germany, E-mail:~\{jing.han,~schuller\}@informatik.uni-augsburg.de \\
         4.\,Educational Physiology Laboratory, The University of Tokyo, Tokyo 113-0033, Japan, E-mail:~qian@p.u-tokyo.ac.jp
}

\begin{abstract}
Acoustic Event Classification (AEC) has become a significant task for machines to perceive the surrounding auditory scene. However, extracting effective representations that capture the underlying characteristics of the acoustic events is still challenging. Previous methods mainly focused on designing the audio features in a `hand-crafted' manner. Interestingly, data-learnt features have been recently reported to show better performance. Up to now, these were only considered on the frame-level. In this article, we propose an unsupervised learning framework to learn a vector representation of an audio sequence for AEC. This framework consists of a Recurrent Neural Network (RNN) encoder and a RNN decoder, which respectively transforms the variable-length audio sequence into a fixed-length vector and reconstructs the input sequence on the generated vector. After training the encoder-decoder, we feed the audio sequences to the encoder and then take the learnt vectors as the audio sequence representations. Compared with previous methods, the proposed method can not only deal with the problem of arbitrary-lengths of audio streams, but also learn the salient information of the sequence. Extensive evaluation on a large-size acoustic event database is performed, and the empirical results demonstrate that the learnt audio sequence representation yields a significant performance improvement by a large margin compared with other state-of-the-art hand-crafted sequence features for AEC. 
\end{abstract}


\end{frontmatter}


\section{Introduction}
\label{sec:introduction}

{{\em Acoustic Event Classification} (AEC) plays an essential role in enabling the environmental awareness for intelligent machines, and has recently attracted considerable attention~(\cite{Chu08-Unstructured,Stowell15-Detection,Ye15-Acoustic,Phan16-Label}).} It can be referred to the field of \emph{Computational Auditory Scene Analysis} (CASA)~(\cite{wang2006computational}), which has an assumption that the audio recordings can be automatically categorised by the location where they were recorded. It has a close relationship to another well-documented task, \emph{Acoustic Event Detection} (AED)~(\cite{phan2014random,xia2019multi}), where the latter focuses on automatically finding accurate onset and offset of an acoustic event (e.\,g., a car passing) in a given duration of audio recording. In many previous studies, the two tasks are jointly implemented in a system that, the audio recordings were segmented by AED then classified into different groups in the scheme of the AEC task. The recent trend including this proposed work prefers a holistic AEC system which does not need a segmentation step in the AEC task.

One central goal of AEC is to extract discriminative representations that are robust enough to capture the acoustic event content. 
In the past decade, many efforts have been reported towards this direction. For example, following the success in speech recognition, Mel-Frequency Cepstral Coefficients (MFCC) have been applied as the most dominant feature type for AEC~(\cite{Ma06-Acoustic,Giannoulis13-Detection}). 
However, unlike speech recognition, AEC highly relies on longer temporal information to make a decision~(\cite{Chu09-Environmental}). In this regard, spectro-temporal features were introduced to capture the event modulation patterns in both time and frequency domains~(\cite{Schroeder15-Spectro,Rakotomamonjy15-Histogram,Ren17-Sound}). 
For instance, a filterbank of two-dimensional Gabor functions was used to decompose the the spectro--temporal power density into multiple components~(\cite{Schroeder15-Spectro}). Similar work was done in~(\cite{Chu09-Environmental}), where a Gabor dictionary was implemented with atom parameters (\ie scale, time, frequency, and phase) for the Matching Pursuit (MP) decomposition, and the generated MP features have shown their efficiency~(\cite{Chu09-Environmental}). Furthermore, another trend aims to build higher-level features from the spectro-temporal features of acoustic event. In this context, Histogram of Gradients (HOG) representation was investigated to provide the information of the changing direction of spectral power~(\cite{Rakotomamonjy15-Histogram}). Besides, a fusion of acoustic and visual features was proposed for the AEC task in~(\cite{xie2019investigation}). Despite its efficiency, it is claimed that using a fusion of acoustic and visual features increases the computational cost. Before executing feature extraction and classification, it was also found that a segmentation step was needed and helps to improve the model robustness in a noisy scenario~(\cite{mulimani2019segmentation}). Although huge efforts have been made on designing optimal features, AEC still remains a challenging task since the audio contains high variety owing to the complex acoustic environment. For this reason, a solution is to combine a variety of features extracted in either time or frequency domain into a fused high-dimensional space~(\cite{Zhang12-Semi}). The primary assumption is that the classification model can automatically select important features for a specific class, which, however, can be quite challenging during model building. 

Another ignored challenge is that most of the previous works were mainly based on Fourier transformation features, which cannot optimise the Heisenberg-like time--frequency trade-off~(\cite{de1967uncertainty}). Therefore, a multi-resolution analysis should be taken into account. Qian~\emph{et al.} used wavelet-based features in~(\cite{qian2017wavelets}) and found that it can improve the performance of the models built by only using Fourier transformation features. Different from these human hand-crafted features, some other models were built on higher representations learnt via \emph{deep learning}~(\cite{LeCun2015}). Ren~\emph{et al.} introduced scalogram-based representations to the AEC task in~(\cite{ren2017deep,ren2018deep}). Furthermore, an attention-based convolutional neural network (CNN) model was proposed in~(\cite{ren2018attention}). Zhang~\emph{et al.} additionally proposed a fine-resolution CNN (FRCNN), which can make the feature representations be easily customised in various time--frequency resolutions~(\cite{zhang2020acoustic}). Abdoli~\emph{et al.} proposed a one-dimensional CNN based system which can learn representations directly from the audio signal, which outperformed the human hand-crafted features and the 2D CNN based representations~(\cite{abdoli2019end}).

Apart from the above mentioned supervised learning scenarios, deep unsupervised representation learning techniques have achieved tremendous success in machine learning~(\cite{Hinton06-fast,Bengio06-Greedy,Bengio13-Representation}). 
The key idea is to learn more complex abstractions as data representations in the higher layers of artificial deep neural networks from simple features in the lower layers in an unsupervised training fashion. The unsupervised representation learning has begun to be applied to AEC, and has shown its efficiency in state-of-the-art research. 
In~(\cite{McLoughlin15-Robust}), Deep Belief Networks (DBN) were employed for pre-training with unlabelled data. The extracted bottleneck features were then fed into a concatenated softmax layer for final classification. To capture the temporal information, sequential frames within a sliding window were batched as the network input. Similarly, a fully Deep Neural Network (DNN) structure was introduced in~(\cite{Xu16-Fully}), where the raw features in continuum were scaled into one super high-dimensional vector and then considered to be the input for a deep `pyramid' structure. All these unsupervised representation learning researches have advanced the performance of AEC systems significantly. 

However, all these works either attempt to learn high-level representations at the frame-level, as the studies did in the field of speech recognition~(\cite{McLoughlin15-Robust,Lee09-Unsupervised}), or assume that the analysed recordings share a fixed duration~(\cite{Xu16-Fully}. Indeed, many event sounds have a strong temporal domain signature as aforementioned. For instance, the chirping of insects is typically noise-like with a broad and flat spectrum, which makes it hard for a system to distinguish it as a noise or an insect sound within one or several audio frames. Moreover, the acoustic events are often presented in arbitrary lengths, rather than fixed lengths. This renders the work in~(\cite{Xu16-Fully}) infeasible in realistic applications. 
To overcome the raised problems for AEC, we propose an {\em unsupervised sequence representation learning} approach, which employs multilayer Grated Recurrent Unit Recurrent Neural Networks (GRU-RNN) to learn representations of audio sequences. The model consists of a RNN encoder to map an input sequence into a fixed-length vector, and a RNN decoder to reconstruct the input sequence from the generated vector into a sequence-to-sequence learning strategy. Our primary assumption is that the representation captures the sequence information as it integrates a `restoration ability' with the help of the decoder. 

The employed encoder-decoder architecture is inspired by the ones used in natural language processing~(\cite{Sutskever14-Sequence}), where the architecture was used for, for example, translating sentences from one language to another~(\cite{Sutskever14-Sequence,Bahdanau14-Neural,Luong15-Effective}), or predicting the next sentence from previous ones~(\cite{Shang15-Neural}). 
Significantly differing from these works, the essential idea of the proposed framework in this article aims to learn a {\em vector representation} of a sequence with an arbitrary length. The learnt representations can then be utilised for pattern recognition by any classification models.

The proposed approach is partially motivated by the work in~(\cite{Srivastava15-Unsupervised}), where a Long Short-Term Memory (LSTM) encoder-decoder architecture was employed for video reconstruction and future prediction. In addition, it relates to~(\cite{Dai15-Semi}) as well, where the LSTM encoder-decoder was utilised for initialising the neural networks and further improving their generalisation capability. The proposed approach, however, is the attempt to obtain a vector representation in a purely unsupervised learning procedure. 

The major contributions of this article mainly include:
 i) We propose an unsupervised learning framework to extract high-level audio sequence representations via a GRU-RNN encoder-decoder for AEC. Compared with previous works, this framework not only can deal with flexible-length audio recordings.  More importantly, it also holds the potential to distil the inherent event characteristics embedded in the audio sequences through infinite unlabelled data, which are ubiquitous and cheep to obtain due to popularity of digital devices.  
 ii) We evaluate the performance of the learnt sequence representations on a large-scale acoustic event database.  The results demonstrate the high effectiveness and robustness of the learnt representations.
iii) The proposed method cannot only benefit the future work in AEC system design, but also other relevant studies in audio or multi-modality related applications.

\section{Related Work}
\label{sec:relatedWork}

There are two dominant methods to represent the audio sequence for AEC. The first method is likely inspired by  speech recognition technology, in which the whole sequence is represented by sequential Low-Level Descriptors (LLDs) (\eg MFCCs) frame by frame. Then, it uses generative models to estimate the joint probability distribution of features and labels to arrive at a final judgment~(\cite{Stowell15-Detection}), or uses discriminative models like by a Support Vector Machine (SVM) to predict the frames successively then voting for a final decision~(\cite{McLoughlin15-Robust}). While the sequence temporal information is going to be utilised as mentioned above, they are still far from being well-explored. 
The second method intends to expand all descriptors and concatenates them into a long vector, and then feeds the vector into a model for discriminative training and evaluation~(\cite{Xu16-Fully}). This method simply assumes that all audio recordings have a fixed length. Also, this method possibly results in a curse of dimension issue when the recording duration increases.
Rather than straightforwardly using the sequential frame-wise LLDs, recent promising methods are more in favour of the sequence-wise {\em statistic} features. 
These methods show the ability to handle the {\em arbitrary-length} recordings, and map them into fixed-dimensional vector representations. 

One efficient method is the {\it Bag-of-Audio-Words} (BoAW)~(\cite{Aucouturier07-bag,Lu14-Sparse}). It uses a codebook of acoustic words (\ie frame-level LLDs) that are randomly selected or generated via a clustering method (\ie $k$-means) on the training set, to quantise the frame-wise LLDs. Then, a histogram of the occurrences of each word in the dictionary is built over the whole sequence, and regarded as the sequence representation.
Another popular method is based on {\it functionals} (\eg mean, standard deviation, skewness, kurtosis, maximum, minimum), which are applied to each of the LLD contours to extract the statistic information over the whole sequence~(\cite{Zhang12-Semi}). 
However, all of these features for audio sequence are still hand-crafted. 

Recently, some pre-trained models are proposed to extract to extract good embeddings from audio, e.\,g., the OpenL3\footnote{https://openl3.readthedocs.io/en/latest/index.html}. These models were pre-trained by large amount of audio data and can be used to train shallow classifiers with limited size of data~\cite{cramer2019look}. We believe this contribution can significantly help overcome the (labelled) data scarcity issue for specific audio classification task. It is noted that building such efficient pre-trained models still need large amount of labelled data and huge computational resources. Additionally, transfer learning technologies are quite dependent on empirical knowledge that can make the pre-trained models work well. In this article, we propose to learn the audio sequence representation in an unsupervised way for the application of AEC. Although a related work has been done in~(\cite{Chung16-Audio}), it is mainly focused on  word-level audio for spoken term detection. To the best of our knowledge, this is the first effort in this direction towards modelling the long audio sequence for a classification task. 

\section{Unsupervised Learning of Audio Sequence Representations}
\label{sec:methodology}
In this article, we are interested in evaluating the performance of a RNN-based sequence-to-sequence encoder-decoder approach for AEC. Before an empirical evaluation, we first describe the proposed method in this section. 

\subsection{Grated Recurrent Unit}
To implement the RNN encoder-decoder, we select the Grated Recurrent Unit (GRU) as the recurrent hidden unit of our RNNs, which was initially proposed by Cho et al.~(\cite{Cho14-properties}). Analogous to the LSTM unit, this recurrent unit can also capture the long-term dependencies in sequence-based tasks and can well address the vanishing gradient problem~(\cite{Chung14-Empirical}). Hence, GRU is often regarded as an alternative to LSTM units. However, the GRU has fewer parameters since it does not have a separate memory cell nor an output gate, which results in a faster training process and less-data demand for generalisation. Besides, many experiments have shown that the GRU performs competitive to or slightly better than the LSTM unit in most tasks~(\cite{Chung14-Empirical,Jozefowicz15-empirical}). 

\begin{figure}[!t]
\centering
%
%
%
%
%

\tikzstyle{object} = [circle, draw, text width=1cm, minimum height=0.5cm, rounded corners, font=\normalsize]
\tikzstyle{action} = [draw, circle,  text width=0.5cm,font=\small]
\tikzstyle{symbol} = [text centered,font=\normalsize]

\tikzstyle{arrow0} = [color=black!80,->, line width=0.5pt]
\tikzstyle{arrow1} = [color=red,->, line width=0.8pt]
\tikzstyle{arrow2} = [color=blue,->, line width=0.8pt]

\tikzset{every node/.style={inner sep=-3pt,minimum height=0.2cm, minimum width=0.2cm,text centered,font=\scriptsize}}

\tikzstyle{vecArrow} = [line width=1.4pt,gray, decoration={markings, mark=at position
   1 with {\arrow[line width=1.4pt, gray]{open triangle 60}}},
   double distance=3pt, shorten >= 7pt,
   preaction = {decorate}]

\makebox[\linewidth]{%
\begin{tikzpicture}[->,>=stealth,node distance = 1cm,auto]

  \node [object](ho) {$h_{t-1}$}; 
  \node [object,left of=ho,node distance = 2cm](hc) {$\tilde{h}_{t}$}; 
  \node [object,right of=ho,node distance = 2cm](z) {$z_t$};
  \node [object,below of=ho,node distance = 1.3cm](r) {$r_t$};
  \node [object,below of=r,node distance = 1.3cm, pin={[pin edge={<-, >=stealth,line width=0.5pt,color=black!70}]south:}](x) {$\mb{x}_t$};
  \node [object,above of=ho,node distance = 1.8cm](y) {$h_t/y_t$};
  \node [symbol,left of=ho](od){$\odot$}; 
  \node [symbol,above of=ho](op){$\oplus$}; 

  \node [symbol,left of=hc](bh) {${b}_h$};
  \node [symbol,right of=z](bz) {${b}_z$};
  \node [symbol,right of=r](br) {${b}_r$};
  
  \node [symbol,below right=.3cm and .3cm of x](in) {input};
  \node [symbol,right of=y,node distance = 1cm](out) {output};

  \draw [arrow0] (x) -| node[above left=.5cm and .1cm]{$W_{xh}$}(hc.south);
  \draw [arrow0] (x) -- node[right=.15cm]{$W_{xr}$}(r.south);
  \draw [arrow0] (x) -| node[above right=.5cm and .1cm]{$W_{xz}$}(z.south);

  \draw [arrow0] (ho) -- (od);
  \draw [arrow0] (r) -| (od);
  \draw [arrow0] (od) -- node[above=.10cm]{$W_{hh}$}(hc);
  \draw [arrow0] (bh) -- (hc);
  
  \draw [arrow0] (ho) -- node[above=.10cm]{$W_{hz}$}(z);
  \draw [arrow0] (ho) -- node[right=.15cm]{$W_{hr}$}(r);
  \draw [arrow0] (ho) -- node[right=.15cm]{$1-z_t$}(op);
  \draw [arrow0] (hc.north) |- node[above right=.0cm and 1cm]{$z_t$}(op);
  \draw [arrow0] (op) -- (y);
  \draw [arrow0] (br) -- (r);
  \draw [arrow0] (bz) -- (z);
  
%

\end{tikzpicture}
}

\caption{Illustration of Gated Recurrent Unit.}
\label{fig:gru}
\end{figure}
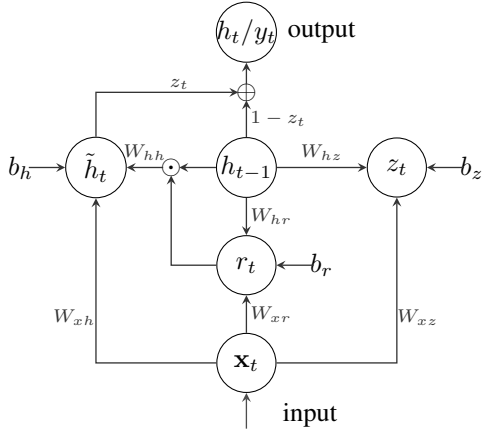

The typical structure of a GRU is depicted in Fig.~\ref{fig:gru}, which consists of a reset gate $r$, an update gate $z$, an activation $h$, and a candidate activation $\tilde{h}$.
Mathematically, let $\mb{x}=(\mb{x}_1,\mb{x}_2, \ldots,\mb{x}_T)$ be an input audio sequence, where $\mb{x}_t\in \Re^d$ is in a $d$ dimension feature space (\eg MFCC). The activation ${h}_t^j$ of the $j$-th GRU at time $t$ is updated by the previous activation ${h}_{t-1}^j$ and the candidate activation $\tilde{{h}}_t^j$, that is 
\begin{equation}\label{eq:1}
 h_t^j=(1-z_t^j)h_{t-1}^j+z_t^j\tilde{h}_t^j.
\end{equation}
The update gate $z_t$ is calculated by 
\begin{equation}
 {z}_t^j=\mathrm{sigm}(W_{xz}\mb{x}_t+W_{hz}{\mb{h}_{t-1}}+\mb{b}_z)^j,
 \end{equation}
where $W$ denotes the weights matrix and $\mb{b}$ stands for the bias vector. 
The update gate $z_t^j$ is used for deciding how much the activation $h_t^j$ is to be updated with a new activation $\tilde{h}_t^j$. Thus, when $z_t^j$ is close to zero, the hidden state almost keeps unchanged in the next time-step. Opposed to this, when $z_t^j$ is close to one, the hidden state will  be overwritten by a new version. In this way, it is expected to maintain any important feature owing to the update gate of the GRU. 

The candidate activation $\tilde{h}_t^j$ is computed mainly by considering the input $\mb{x}_t$, the reset gate $\mb{r}_t$, and the previous time-step hidden activation $\mb{h}_{t-1}$, as follows 
\begin{equation}
 \tilde{{h}}_t^j=\mathrm{tanh}(W_{xh}\mb{x}_t+W_{hh}(\mb{r}_t\odot\mb{h}_{t-1})+\mb{b}_h)^j,
\end{equation}
where $\odot$ is an element-wise multiplication, and $\mb{r}_t$ is a set of reset gates. Here, the reset gate ${r}_t^j$ decides on how much the previous activation ${h}_{t-1}^j$ impacts the candidate activation $\tilde{{h}}_t^j$. Only when ${r}_t$ equal zero, the candidate activation will  be overwritten by the current inputs. Similar to the update gate, the $j$-th reset gate is computed by
\begin{equation}\label{eq:4}
 {r}_t^j=\mathrm{sigm}(W_{xr}\mb{x}_t+W_{hr}\mb{h}_{t-1}+\mb{b}_r)^j. 
\end{equation}

\subsection{Audio Sequence Representation Learning} 

\begin{figure}[!t]
\centering
\includegraphics[trim=2.2cm 0cm 2cm 0cm, width=2.8in,height=1.2in]{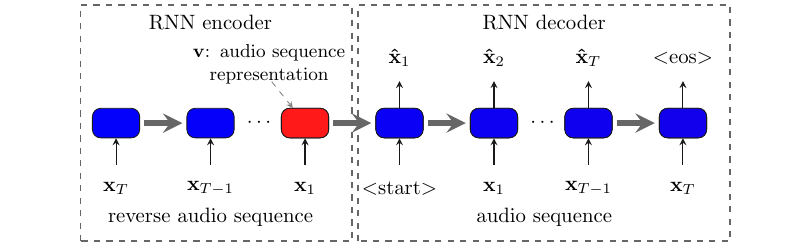}
\caption{Unsupervised framework for learning of audio sequence representations with a sequence-to-sequence Recurrent Neural Network (RNN) encoder-decoder.} 
\label{fig:rnnAutoencoder}
\end{figure}

The proposed unsupervised representation learning framework of audio sequences is illustrated in Fig.~\ref{fig:rnnAutoencoder}, which comprises a GRU-RNN {\em encoder} and a GRU-RNN {\em decoder}.
The primary objective of this framework is to transform an arbitrary-length audio segment, give\re{n} as a sequence of feature vectors  $\mb{x}=(\mb{x}_1,\mb{x}_2, \ldots,\mb{x}_T)$, into {\em one fixed-length} vector representation $\mb{v}$. 

Specifically, the RNN encoder reads the acoustic feature $\mb{x}_t$ sequentially and {\em reversely} as done in~(\cite{Sutskever14-Sequence}), and the hidden state $\mb{h}_t$ is updated accordingly by 
\begin{equation}
 \mb{h}_t=f(\mb{x}_t,\mb{h}_{t+1},\mb{z}_{t},\mb{r}_{t})
\end{equation}
where $f$ denotes the GRU activation function as introduced in the above section.
After the last acoustic feature $\mb{x}_1$ has been read and processed, the hidden state $\mb{h}_1$ of the RNN encoder can be viewed as the learnt vector representation $\mb{v}$ of the whole input sequence.

The decoder aims to reconstruct the input sequence of the encoder in a {\em `normal'} direction, as $\mb{\hat{x}}=(\mb{\hat{x}}_1, \mb{\hat{x}}_{2}, \ldots, \mb{\hat{x}}_{T})$. To do this, the last hidden state $\mb{h}_1$ of the encoder is copied to the decoder as its initial hidden state, \ie $\mb{\hat{h}}_1=\mb{h}_1$.

Then, the decoder predicts the feature vector $\mb{{\hat{x}}_{t}}$ by given its hidden state $\mb{\hat{h}}_{t}$, update gate $\mb{\hat{z}}_{t-1}$, reset gate $\mb{\hat{r}}_{t-1}$, and its input $\mb{x}_{t-1}$, that is,
\begin{equation}
 \mb{{\hat{x}}}_t = g(\mb{x}_{t-1},\mb{\hat{h}}_{t},\mb{\hat{z}}_{t-1},\mb{\hat{r}}_{t-1}),
\end{equation}
where $g$ is the GRU activation function as well. 
Note that, rather than using the previously predicted feature sequence, we utilise the original feature sequence as the decoder input, which is motivated by the finding of the work~(\cite{Bengio15-Scheduled}). That is, the original feature sequence is helpful in improving the model robustness to its own errors when training. 

The RNN encoder and decoder are jointly trained by minimising the reconstruction error, measured by the averaged Mean Square Error (MSE): 
\begin{equation}
 \frac{1}{T}\sum_{t=1}^T\parallel\mb{x}_t-\mb{\hat{x}}_t\parallel^2.
\end{equation}
The whole training process is carried out in a fully {\em unsupervised} manner since label information is not required at all. 

Finally, when the audio sequences are fed into the pre-trained encoder-decoder, and the last hidden state of the encoder for each audio sequence will be viewed as its fixed-dimensional vector representation $\mb{v}$. Since this vector is able to reconstruct itself by the RNN decoder, we believe that such a vector contains the whole sequence information in a compressed way.

\section{Experiments and Results}
\label{sec:experiments}

In this section, we are devoted to estimating the effectiveness and the robustness of the proposed framework for learning audio sequence representations. Extensive experiments are conducted on a large-size acoustic event database, and the empirical results are compared with other state-of-the-art baselines. 

\begin{table}[t]
\centering
\caption{Quantitative description of Findsounds2016.}
\vspace{.1cm}
\begin{threeparttable}
\begin{tabular}{lrrrr} 
\toprule
\bf category & \bf \# class & \bf \# segment & \bf duration\\ 
\midrule
  Animals & 67 & 1\,998 & 1h 53m \\ 
  Birds & 102 &  1\,766 & 1h 53m \\
  Household & 53 & 2\,097 & 1h 27m \\
  Insects & 7 & 235 & 16m \\
  Mayhem & 35 & 1\,471 & 50m \\
  Miscellaneous  & 70 & 2\,628 & 1h 45m \\
  Musical Instruments & 57 & 4\,112 & 3h 35m \\
  Nature & 18 & 754 & 1h 3m \\
  Office & 18 & 1\,188 & 50m \\
  People & 45 & 2\,165 & 1h 44m \\
  Sports Recreation & 22 & 266 & 9m \\
  Tools & 21 & 296 & 18m \\
  TV Movies & 22 & 645 & 24m \\
  Vehicles & 33 &  1\,714 & 2h 9m \\ 
\midrule
\bf{Total} & \bf 570 & \bf 21\,335 & \bf 18h 23m \\ 
\bottomrule
\end{tabular}
\end{threeparttable}
\label{tab:findsounds}
\vspace{-.0cm}
\end{table}

\subsection{Database Description}
\begin{figure}[t]
\centering
\resizebox{0.45\textwidth}{2.7in}{\input{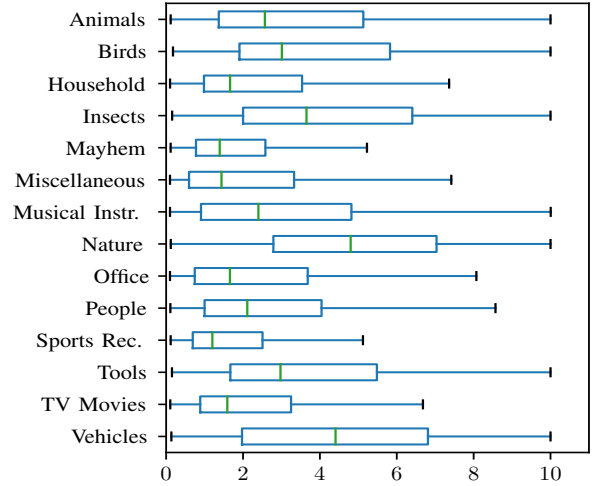}}
\caption{Duration distribution of audio segments for each acoustic-event category over the whole Findsounds2016 database.}
\label{fig:dataDistribution}
\end{figure}

The database selected for our experiments -- {\em Findsounds2016} -- is supposed to be a large publicly available databases for the AEC research when conducting the experiments~(\cite{Piczak15-ESC:}). It was collected from the website of `www.findsounds.com', which provides a comprehensive set of event-annotated audio recordings from real environments, reaching from nature (\eg nature and animals) over human beings (\eg people) to manufactured articles (\eg musical instruments and vehicles). Specifically, we discarded two categories (\ie Holidays and Noisemakers) from the original dataset due to the sample-overlapping with other categories, resulting in a final set of 14 common acoustic-event categories. Each category further includes a number of classes (subsets), giving rise to a total of 570 classes and 21\,335 independent audio segments, with a total duration of more than 18 hours. 
More details on the number of segments and recording time per category are summarised in Table~\ref{tab:findsounds}. 
The averaged duration over all audio segments is 3.1\,s with a maximum and a minimum of 10.0\,s and 0.1\,s, respectively. 
{In detail, Fig.~\ref{fig:dataDistribution} illustrates the duration distribution for each acoustic-event category over the whole database. Obviously, these duration distributions are highly overlapped and mainly range from one to six seconds.}
Moreover, owing to the diversity of the audio formats in the original dataset retrieved from the web, we converted all audio files into a uniformed format with 16-bit encoding, mono-channel, and 16\,kHz sampling rate. It is noted that the majority samples of this dataset are monophonic audio. Only a limited or negligible number of overlapped sounds exist.

For training the back-end classifier, each subset of the Findsounds2016 database was equally and sequentially partitioned into training set (7\,312 instances), test set (7\,106 instances), and validation set (6\,917 instances). In addition, we always upsampled the training set to alleviate the unbalanced class-distribution problem, by randomly repeating the samples in less dominated classes several times with the random seed zero.

\subsection{Experimental Setup} 

\begin{figure*}[t]
\centering
\subfigure[{SVMs}]{
\includegraphics[width=0.48\textwidth]{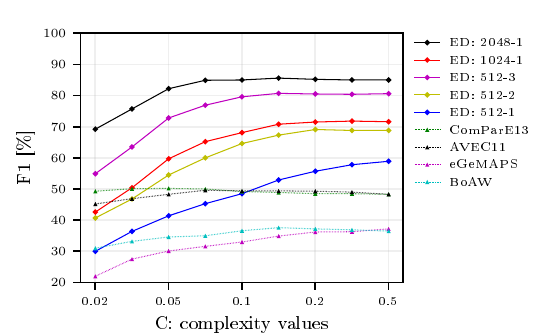}
}
\subfigure[{GRU-RNNs}]{
\includegraphics[width=0.48\textwidth]{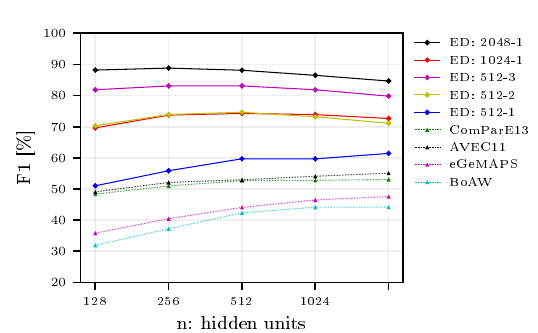}
} 
\vspace{0cm}
\caption{Performance comparison (F1-measure) between the {\em learnt audio sequence representations} via a variety of RNN Encoder-Decoders (ED) and four {\em hand-crafted features} on the {\em validation} set of Findsounds2016. Performance was evaluated by (a) the SVMs with various complexity values, $C$, or (b) the GRU-RNNs with various numbers of hidden units, $n$.}
\label{fig:resultsValid}
\vspace{0cm}
\end{figure*}

For training RNN encoder-decoders to learn the audio sequence representations, we theoretically could feed the raw signals into the network directly. However, the long sequence-length leads to a high requirement of computational resource. As MFCCs have been repeatedly verified to be the efficient features for most acoustic recognition tasks and we have limited computational resource, we extracted 13 MFCCs (including one logarithmic energy) per frame using a window size of 60\,ms at a step size of 60\,ms. Compared with the conventional parameters for extracting MFCCs (\ie window size: 25\,ms, step size: 10\,ms), the ones we selected due to i) the audio pattern normally changes slower than the speech pattern. In this case, within a longer frame window size, the audio signal is still considered to be stationary; ii) a longer frame window size is helpful to extract rich low fundamental frequency information, which fits for audio analysis; iii) a longer frame step size compared to one used for speech can reduce the extracted number of frames and consequently significantly speed up the network training process.

In this case, the longest sequence of Findsounds2016 has 167 MFCC feature vectors. Finally, all the extracted features were standardised by the means and variations of the training set.  

To accelerate the  RNN encoder-decoder training process, we used a mini batch of 64 sequences as network input. In this case, we padded zeros at the end of each sequence to force them equilong. The padded zeros, however, are ignored when calculating the reconstruction errors (\ie training loss) in the training process by setting their weights to zero. Further, to control the learning process, we checked the training loss after running every 500 batches. To update the network weights, we employed the classic Stochastic Gradient Decent (SGD) with an initial learning rate of 0.7. This value dynamically reduced with a decay factor of 0.99 when the training loss was not improved any more over the previous three checking points. Additionally, a gradient norm clipping operation was performed with a clipping ratio of 5 to handle the gradient blowup problem. The whole learning process was stopped once there was no training loss improvement over 20 successive checking points. 

To assess the discrimination and the robustness of the learnt audio sequence representations via the pre-trained RNN encoder-decoder, we further adopted nowadays two of the most frequently used classification models. One of them is the SVM trained with the sequential minimal optimisation algorithm. The complexity value of $C$ was optimised in \{0.01, 0.02, 0.05, 0.1, 0.2, 0.5, 1, 2, 5\} on the validation set. Another one is the GRU-RNN with one hidden layer, while the number of hidden units was optimised in \{128, 256, 512, 1024\} on the validation set.  Additionally, the GRU-RNNs were trained with Adam SGD with an initial learning rate of $10^{-4}$, to which an exponential decay was applied at every $10^{4}$ steps with a decay rate of 0.96. Further, the gradient norm clipping ratio was set to 1.2, and the batch size was set to 128. For equal comparison, the training processes of all networks were stopped at the 500th epoch. 

To measure the system performance, we utilised {\em F1-measure} (F1) as a primary metric, 
{mainly due to the facts that i) F1 provides an overview performance in a multi-class setting as it is calculated by the harmonic mean of unweighted precision and recall; ii) F1 is among the most widely used evaluation metrics in AEC, for example, in a series of challenges of Detection and Classification of Acoustic Scenes and Events (DCASE)~(\cite{Stowell15-Detection,Mesaros16-TUT}).}
Additionally, we took the {\em Unweighted Accuracy} (UA, or unweighted recall) as a complementary metric. It is obtained by the sum of the accuracies over all classes divided by the number of classes. Thus, UA also well indicates the system performance in a class-unbalanced task.

Further, unless stated otherwise, a one-side $z$-test was undertaken to evaluate the statistical significance of performance improvement.

\subsection{Compared Features of Audio Sequence}
To verify the effectiveness of the learnt representation of audio sequence, we selected one BoAW feature set and three functional-based feature sets for comparison. All these feature sets are widely used for AEC or related acoustic tasks (\eg emotion) nowadays. A brief description of the four feature sets is as follows: 
\begin{itemize}
\item {\it BoAW} feature set: The codebook includes 2\,048 audio words. Each frame of the sequence is then assigned to the nearest 256 audio words. Afterwards, a normalised histogram is applied to convert the word occurrence accounts into a fixed-length vector~(\cite{Lu14-Sparse}). 
\item the extended Geneva Minimalistic Acoustic Parameter Set ({\it eGeMAPS}): It consists of 88 important acoustic attributes, which were selected by extensive experiments on acoustic pattern classification tasks~(\cite{Eyben16-Real}). 
\item the 2011 Audio-Visual Emotion recognition Challenge \\({\it AVEC11}) feature set: It contains 1\,941 attributes and was used in~(\cite{Zhang12-Semi}) for AEC. 
\item the INTERSPEECH 2013 Computational Paralingusitics ChallengE ({\it ComParE13}) feature set: It includes a large-scale acoustic attributes up to 6\,373~(\cite{Eyben16-Real}). 
\end{itemize}

\subsection{Results}

To evaluate the robustness of the proposed framework, we constructed the RNN encoder-decoders in several structures, mainly towards a {\em deep} or a {\em wide} direction. To assess the deeper networks, we fixed the number of hidden units per layer as 512, and then set the hidden layers to one, two, or three, resulting in three RNN encoder-decoders in different depths. To assess the wider networks, we fixed the depth of hidden layer as one, but set the number of hidden units to 512, 1\,024, or 2\,048, leading to additional two RNN encoder-decoders in different widths. Note that the RNN encoders and corresponding decoders always share the same structures.

Fig.~\ref{fig:resultsValid} illustrates the performance  of the learnt representations obtained by diverse RNN encoder-decoders, as well as four conventional feature sets based on BoAW or functionals (\ie eGeMAPS, AVEC11, and ComParE13). 
The performance was estimated on the  validation set of Findsounds2016 for 14 acoustic-event categories.

Specifically, Fig.~\ref{fig:resultsValid} (a) depicts the feature performance when taking the SVMs as discriminative models. 
From this figure, one can obviously observe that the results delivered by the learnt representations are remarkably higher than the other four state-of-the-art baselines. The best result is achieved at 85.6\,\% of F1 by using the representations learnt by the RNN encoder-decoder with one hidden layer of 2\,048 hidden units (ED: 2048-1). This result is almost double of the best baseline achieved by using ComParE13 or AVEC11 feature set (\ie 50.2\,\% of F1). 

Further, when increasing the depth of the neural networks from one to two and three, one can see a steady and significant performance improvement. Similarly, when extending the width of the neural networks from 512 to 1\,024 and 2\,048, again huge performance improvement is obtained. This indicates that appropriately increasing the complexity of the sequence-to-sequence model, either in a deep way or in a wide way, can notably improve the effectiveness of the learnt representations. 

\begin{table}[t]
\centering
\caption{Performance comparison (F1 and UA) between the {\em learnt audio sequence representations} via a variety of RNN Encoder-Decoders (ED) and four  {\em hand-crafted features} on the {\em test} set of Findsounds2016. Performance was evaluated by both SVMs and GRU-RNNs.} 
\vspace{.15cm}
\begin{threeparttable}
\begin{tabular}{c|cc|cc} 
 \toprule
  $[\%]$ & \multicolumn{2}{c|}{\bf SVMs} & \multicolumn{2}{c}{\bf GRU-RNNs} \\ 
 feature types & F1 & UA & F1 & UA \\
 \midrule
 BoAW &41.9 &35.3 &44.4 &39.5 \\ 
 eGeMAPS &36.4 &34.9 &47.6 &41.4 \\ 
 AVEC11 &50.4 &42.8 &54.0 &48.7 \\ 
 ComParE13 &49.7 &43.6 &53.2 &46.2 \\ 
 \midrule
 ED: 512-1 &58.1 &52.9 &61.1 &54.9 \\
 ED: 512-2 &68.4 &63.4 &71.8 &67.4 \\
 ED: 512-3 &80.6 &76.6 &80.5 &78.4 \\ 
 ED: 1024-1 &72.0 &65.8 &72.6 &70.0 \\
 ED: 2048-1 &\bf85.2 &\bf80.4 &\bf89.0 &\bf87.6 \\
 \bottomrule
\end{tabular}
\end{threeparttable}
\label{tab:resultsTest}
\end{table}

Similar observations can be found in Fig.~\ref{fig:resultsValid} (b), where GRU-RNNs were employed as discriminative models. Generally speaking, however, GRU-RNNs yield better performance than SVM in all cases. The best result further rockets to 88.8\,\% of F1. Additionally, an interesting observation can be seen that the learnt representations performs better when using relatively simple networks for classification, yet the hand-crafted features incline to choose the relative complex networks for classification in order to get better results. This indicates that the learnt representations is easier to be learnt by a simple machine learning model than the selected hand-crafted features. 

\begin{figure}
\centering
\resizebox{0.48\textwidth}{2.15in}{\input{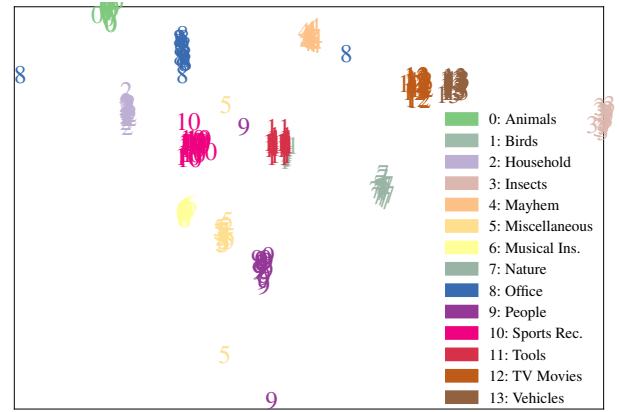}}
\caption{Visualisation of the audio sequence representations learnt from the RNN encoder-decoder. 20 samples per class are randomly selected through the whole dataset and projected into the leading two discriminant directions found by Linear Discriminant Analysis (LDA). Each sample is remarked by the category number (0$\sim$13) it belongs to.}
\label{fig:representVisual}
\end{figure}

We further evaluated the learnt representations on the test set by employing both SVMs and GRU-RNNs for classification with the best parameter settings optimised on the validation set. Table~\ref{tab:resultsTest} displays the corresponding results in terms of F1 and UA. Consistently, the RNN encoder-decoder with 2048 hidden units offers the most efficient features, contributing to 85.2\,\% of F1 and 80.4\,\% of UA by means of SVMs, and 89.0\,\% of F1 and 87.6\,\% of UA by means of GRU-RNNs. Compared with the best baseline, they provide absolute gains as high as 35.0\,\% of F1 and 38.9\,\% of UA. 
To further investigate the effectiveness of the learnt representation, we randomly selected 20 samples from each categories and projected them into the leading two discriminant directions found by Linear Discriminant Analysis (LDA). The visualisation of the audio sequence representations is displayed in Fig.~\ref{fig:representVisual}. 
Notably, the samples belong to different categories are strongly discriminative, which reasonably results in a high prediction accuracy. 

\begin{figure}
\centering
\resizebox{0.48\textwidth}{2.8in}{\includegraphics[width=0.48\textwidth]{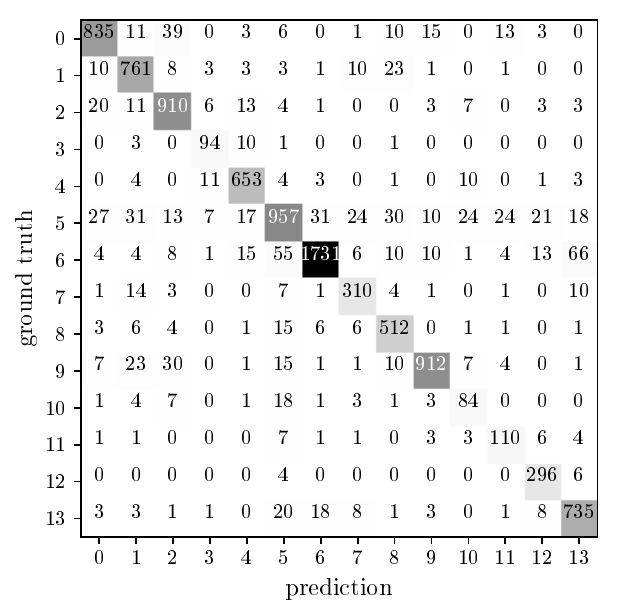}}
\caption{Normalised confusion matrix ($\times 10^{-4}$) of predictions on the test set obtained by the best GRU-RNN classification model (one hidden layer with 128 hidden units). The labels from 0 to 13 sequentially indicate the categories from `Animals' to `Vehicles' as listed in Table~\ref{tab:findsounds} within the same order.}
\label{fig:confusionMatrix}
\end{figure}

To intuitively demonstrate the best performance we achieved by using the GRU-RNN based classifier with one hidden layer and 128 hidden units, Fig.~\ref{fig:confusionMatrix} illustrates the prediction confusion matrix on the test set, which is obtained by using the vector representations learnt by the RNN encoder-decoder comprised of one hidden layer with 2\,048 hidden units. Generally speaking, the acoustic segments represented by the proposed vectors can be well distinguished into corresponding categories. In more detail, one can notice that the category of `Miscellaneous' (labelled as 5 in Fig.~\ref{fig:confusionMatrix}) is relatively easier to be misclassified into the others, which keeps in line with the fact that its contents include many quite similar acoustic events to the other categories.

In addition, we performed the same experiments on the acoustic-event classes. Rather than utilising the whole 570 classes, we discarded those classes having extremely sparse samples (fewer than 20). This leads to a subset of 229 selected classes, and a slightly smaller training set (6\,277), test set (6\,202), and validation set (6\,122). Table~\ref{tab:resultsSubclasses} shows the corresponding results for various features or representations. Interestingly, the learnt representations consistently outperform the frequently used feature sets, and yield the highest F1 and UA of 47.7\,\% and 39.0\,\%, respectively, for 229 types of acoustic events. 

\begin{table}[t]
\centering
\caption{Performance comparison (F1 and UA) for classifying {\bf 229} classes of acoustic events between the {\em learnt audio sequence representations} via a variety of RNN Encoder-Decoders (ED) and four {\em hand-crafted features} on the {\em test} set of Findsounds2016. Performance was evaluated by both SVMs and GRU-RNNs.} 
\vspace{.15cm}
\begin{threeparttable}
\begin{tabular}{c|cc|cc} 
 \toprule
  $[\%]$ & \multicolumn{2}{c|}{\bf SVMs} & \multicolumn{2}{c}{\bf GRU-RNNs} \\ 
 feature types & F1 & UA & F1 & UA \\
 \midrule
 BoAW &18.5 &17.8 &17.7 &17.5 \\ 
 eGeMAPS &18.2&20.1&21.8&20.9 \\ 
 AVEC11 &26.8&23.6&20.0&18.8 \\ 
 ComParE13 &27.1&23.8&23.1&21.5 \\ 
 \midrule
 ED: 512-1 &19.9 &20.1 &25.6&23.0 \\
 ED: 512-2 &29.0 &26.4 &31.5&27.3 \\
 ED: 512-3 &34.6 &\bf32.0 &43.2&36.5 \\
 ED: 1024-1 &24.5 &23.8 &32.6&28.5 \\
 ED: 2048-1 &\bf35.1 &30.8 &\bf47.7&\bf39.0 \\
 \bottomrule
\end{tabular}
\end{threeparttable}
\label{tab:resultsSubclasses}
\end{table}

In future work, one needs to consider implementing our method in an unsupervised learning paradigm, which has been demonstrated to be successful in extracting semantic representations from large size data set~(\cite{jansen2018unsupervised}).

\section{Conclusions}
\label{sec:conclusions}

In this article, we proposed an unsupervised framework to learn the essential patterns of acoustic events that are embedded through the whole audio sequence. In this framework, a Recurrent Neural Networks (RNN) based sequence-to-sequence encoder-decoder is used, where the inputs are the sequential and reverse acoustic feature vectors and the targets are their counterparts in normal order. This encoder-decoder is trained without any category information such that it has the huge potential to explore big unlabelled data in the real world. 
We then extracted the bottleneck features as the audio sequence representations for acoustic event classification, and evaluated them through traditional machine learning algorithms. 

This framework can address the audio sequences with arbitrary durations, and compress them into vector representations with a fixed dimension. Since the learnt representation can be well recovered to its original version by the decoder, it is thus supposed to contain the most important sequence information. The effectiveness and robustness of the proposed framework was extensively examined by the experiments on a large dataset, which have raised the state-of-the-art baselines into a significantly high level.  

Encouraged by the achieved results, we will further evaluate our proposed method in a recently released weekly labelled dataset AudioSet~(\cite{Gemmeke17-Audio}). We believe that the proposed learning representation approach is a major breakthrough in the development of the RNN based encoder-decoder models, which could potentially lead to a range of exciting applications way out of our chosen exemplary application. These applications, which highly characterised with sequential patterns via either audio or video signals, include activity detection, emotion recognition, polyphonic sound tagging, and the like.

\section*{Acknowledgment}

This work was partially supported by the Zhejiang Lab's International Talent Fund for Young Professionals (Project HANAMI), China, the JSPS Postdoctoral Fellowship for Research in Japan (ID No.\,P19081) from the Japan Society for the Promotion of Science (JSPS), Japan, the Grants-in-Aid for Scientific Research (No.\,19F19081 and No.\,20H00569) from the Ministry of Education, Culture, Sports, Science and Technology (MEXT), Japan, the European Union's Seventh Framework Programme through the ERC Starting Grant No.\,338164 (iHEARu), and the Horizon 2020 Programme through the Research Innovation Action No.\,645094 (SEWA).

\bibliographystyle{elsarticle-harv}\biboptions{authoryear}
\bibliography{references}







\end{document}